\documentclass{appolb}
\usepackage{graphicx}
\usepackage{authblk}

\begin{document}
\title{Overview of the software architecture and data flow for 
the J-PET tomography device%
\thanks{Presented at ...}%
}
\author{W.~Krzemie\'n$^{c}$, D. Alfs$^{a}$, P.~Bia\l as$^{a}$, E.~Czerwi\'nski$^{a}$, A.~Gajos$^{a}$, B.~G\l owacz$^{b}$, B.~Jasi\'nska$^{b}$, D.~Kami\'nska$^{a}$, G.~Korcyl$^{a}$,  P.~Kowalski$^{f}$, T.~Kozik$^{a}$,  E.~Kubicz$^{a}$, Sz.~Nied\'zwiecki$^{a}$, M.~Pawlik-Nied\'zwiecka$^{a}$, L.~Raczy\'nski$^{f}$, Z.~Rudy$^{a}$, M.~Silarski$^{a}$, A.~Strzelecki$^{a}$, A.~Wieczorek$^{a,d}$, W.~Wi\'slicki$^{f}$, M.~Zieli\'nski$^{a}$, P.~Moskal$^{a}$}

\affil{
       
       $^{a}$Faculty of Physics, Astronomy and Applied Computer Science, Jagiellonian University, 30-348 Cracow, Poland\\
       $^{b}$Department of Nuclear Methods, Institute of Physics, Maria Curie-Sk\l odowska University, 20-031 Lublin, Poland\\
       $^{c}$High Energy Physics Division, National Center for Nuclear Research, 05-400 Otwock-\'Swierk, Poland\\
       $^{d}$Institute of Metallurgy and Materials Science of Polish Academy of Sciences, 30-059 Cracow, Poland\\
       $^{e}$Faculty of Chemistry, Jagiellonian University, 30-060 Cracow, Poland\\
       $^{f}$\'Swierk Computing Center, National Center for Nuclear Research, 05-400 Otwock-\'Swierk, Poland\\
     }

\maketitle
\begin{abstract}
Modern TOF-PET scanner systems require high-speed computing resources for efficient data processing, monitoring and image reconstruction. 
In this article we present  the data flow and 
software architecture for the novel TOF-PET scanner developed by the J-PET collaboration. 
We discuss the data acquisition system, reconstruction framework and image reconstruction software.       
Also, the concept of computing outside hospitals in the remote centers such as \'Swierk Computing 
Centre in Poland is presented.
\end{abstract}
\PACS{PACS numbers come here}
\section{Introduction}
Positron Emission Tomography (PET) is at present one of the most technologically advanced imaging techniques used in medical diagnosis.
It allows for non-invasive tomographic imaging of physiological processes in-vivo.
The gamma quanta pairs given off by a radioactive tracer administered to the patient's body,
are registered in coincidence by the PET scanner detector to reconstruct image of the tracer spatial distribution.
The significant improvement of the image contrast and the faster convergence of image reconstruction procedure can be achieved by applying the Time-of-Flight (TOF) technique~\cite{Karp2008,Kardmas2009} based on the determination of the annihilation point along the Line of Response (LOR) by measuring the time difference between the arrival of the gamma quanta at the detectors.

  The current commercial PET devices use inorganic crystal scintillators for the detection of the gamma quanta. 
In contrast, the J-PET collaboration is developing a prototype PET based on polymer scintillators~\cite{patent_pm,patent_pm2,NovelDetectorSystems,NIM15PM,NIM15LR,NIM14LR,NIM14PM,StripPETconcept,TOFPETDetector}. 
This novel approach exploits the excellent time properties of plastic scintillators, 
which permit a very precise time measurement, making the usage of the TOF technique more effective.
The obtained timing properties allow to extend the J-PET scanner application
to studies in fields such as material science~\cite{x4,x5}, nano-biology~\cite{x3} or to investigation of fundamental symmetries violation in ortho-positronium system~\cite{x6,x7}.

\begin{figure}[h!]
  \centerline{\includegraphics[width=0.8\textwidth]{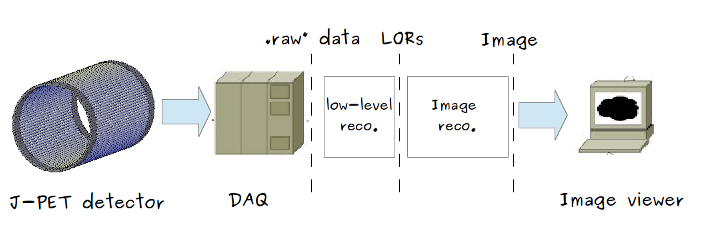}}
\caption{
Scheme of the data flow in the PET image reconstruction. The raw data collected by the Data Acquisition System (DAQ) is then processed by the low-level reconstruction module (low-level reco.) and results in the set of reconstructed lines of response (LORs), which are then sent to the image reconstruction module (Image reco.). The final image can be visualized by the dedicated image viewer or can be exported to DICOM format.
\label{fig:data_flow}
}
\end{figure}

The use of state-of-art detectors together with a dedicated Data Acquisition 
System (DAQ) imposes new requirements for the processing of the data streams, 
for the monitoring, as well as for the reconstruction procedures. 
The current prototype of the TOF-PET scanner being developed by the J-PET 
collaboration will consist of 192 of detection modules with double-sided readout, 
made out of photomultiplier and front-end electronics (FEE). 
The FEE allows to sample
the output time signals at 4 separated voltage levels corresponding 
to 8 samples in total: 
4 for signal rising and 4 for falling edges. 
This sums up to $192 \times 2 \times 8 $ output channels 
per event. 
In addition, the front-end electronics work in the so-called trigger-less mode~\cite{korcyl, korcyl2}, 
storing all incoming events without master-trigger conditions applied. 
This results in a big data flow that needs to be handled and stored efficiently. 
The collected data are processed in several steps (see~Fig.\ref{fig:data_flow}) 
of low- and high-level reconstruction leading to a significant data volume reduction. 
At the same time the information needed to obtain the final image of the human body is preserved.
  The process starts with the collection of the raw data (time and amplitude are digitized by the time-to-digital and analog-to-digital converters). Next,  the data is combined into signals and  translated to hit positions in the individual scintillator modules. Finally, hits in the individual detector bars are combined to form LORs. The set of LORs is then used as an input for the image reconstruction procedure. 
To further speed up the data processing, several parallelization techniques are applied at different stages of computing~\cite{krzemien_parallel}.

  In the next sections, we describe in more details the front-end electronics, reconstruction framework and image reconstruction techniques developed  by the J-PET collaboration. We also present the  distributed computing architecture.
\section{Front-end electronics}
One of the main novelties of the proposed J-PET detector lays in the reconstruction of gamma quanta hit position in the polymer module by performing a very precise time measurement. This  method puts hard requirements on the read-out electronics. 
A typical signal rising time from the polymer scintillator used in the J-PET project is about 0.5 ns what combined with the rising time of a fast photomultiplier (e.g. R4998) of about 0.7 ns, results in a signal rising time of about 1 ns. This value allows to obtain an excellent time resolution but at the same time the read-out electronics must have a much better accuracy to sample those short-time signals e.g. probing a signal with a rising time of about 1.2 ns on 16 levels, would require an accuracy roughly below 75 ps.  
Unfortunately, the existing solutions do not provide the expected time precision. 
Therefore, one of our main efforts was aimed to construct the read-out electronics 
which would fulfil the requirements of the detector and at the same time would keep costs reasonably low. 
The design allows for sampling in the voltage domain of signals with a duration of few
nanoseconds~\cite{palka}. 
The FEE solution is a purely
digital implementation, based solely on a FPGA (Field Programmable Gate Array)
device and a few satellite discrete electronic components.
The input signals are amplified
and split into four paths, each
having an individual threshold
level. The design includes only
DAC chips (LTC2620) for
threshold settings and passive
splitters connected to the FPGA
Low Voltage Differential
Signalling (LVDS) buffers.
Time to digital conversion is
realized in the FPGA and it is
based on low-delay carry-chains
usually used as a part of adders.
The solution allows to probe the
signal in the voltage domain
with an accuracy below 20 ps
($\sigma$). 
The charge of the signal is determined
by the Time-over-Threshold method (ToT), which
is based on the relation between the charge
and the signal width.
An additional advantage of
the FPGA solution is a very low
cost. 
At present, in the
prototype phase, the cost 
together with digitization is
only about 10 Euro per sample.
The read-out electronics permit a multi-threshold  sampling, which probes the signal event waveform with respect to four amplitude thresholds.  
According to the Compressive Sensing Theory~\cite{CStheory1}, the collected data points could be used to reconstruct the full signal shape e.g. by applying the transformation to a sparse representation. 
The information about the shape of the signals is highly correlated with the hit position of the annihilation gamma quantum along the scintillator strip. Thus, with this information a better filtering of coincidence of the two signals and also a more accurate reconstruction of the position are possible~\cite{NIM15PM,NIM14LR,NIM15LR,moskal_neha}.

The FPGA devices are also the core computing nodes of the JPET DAQ,
which  allows for so called continuous or triggerless data taking mode in which the data are collected without any central trigger selection conditions~\cite{korcyl, korcyl2}.

\section{Low- and high-level data processing}
\begin{figure}[h!]
  \centerline{\includegraphics[width=0.8\textwidth]{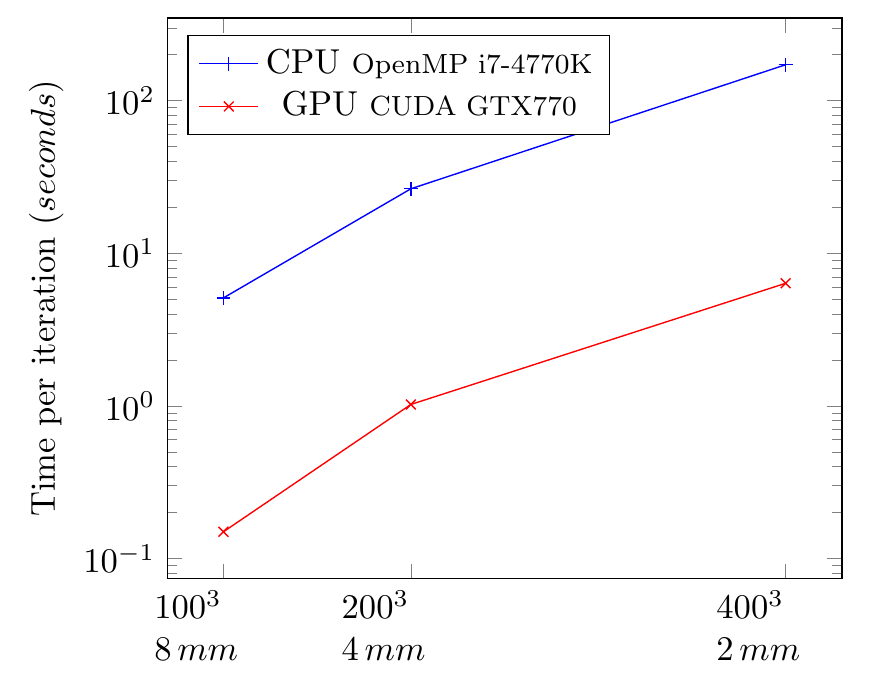}}
\caption{
\label{fig:cpu_gpu}
CPU vs GPU implementation -- single image reconstruction iteration time as a function of target image resolution.
}
\end{figure}

The raw data provided by the front-end electronics is processed in the low-level  reconstruction framework, which serves as a backbone system for various algorithms (e.g. aforementioned position reconstruction algorithms), calibration procedures and to standardize the common operations, e.g: input/output process, access to the detector geometry parameters and more~\cite{krzemien}. 
The framework has been developed in C++ using the  object-oriented approach. It is based on the BOOST~\cite{boost} and ROOT~\cite{root} programming libraries. The framework is used for the off-line analysis,  but also as an on-line module being a part of the steering software system PetController.
The next step in the data processing is the reconstruction of the radioactivity distribution in the patient's body based on the collected LORs. We adopted the most popular  approach based on iterative algorithms derived from Maximum Likelihood Expectation Maximization (MLEM)~\cite{shepp}. The available time-of-flight information is incorporated to improve the accuracy  and the quality of the reconstruction.
  The image reconstruction is one of the most time-consuming parts of the whole data flow. In order to reduce the processing time, parallelization techniques are applied. Currently, a solution based on  a multi-core CPU architecture is implemented~\cite{slomski}. Efforts are made to exploit the processing capability of Graphical Processing Units (GPU). The  efficient image reconstruction using list-mode MLEM algorithm with  approximation kernels was implemented on GPU~\cite{bialas1,bialas2,bialas_GPU}. 
Current J-PET GPU-based reconstruction algorithm is able to provide the full 3D reconstruction image of $200^3$ 2mm voxels in about one minute time, by exploiting the time-of-flight information.
The comparison between CPU and GPU reconstruction time per iteration for the sample Shepp-Logan~\cite{shepp_logan} simulated phantom is presented in~Fig.\ref{fig:cpu_gpu}.


\section{Remote processing and storage}

Apart from the presented computing schemes, 
in which the data processing is performed locally, 
using multi-core CPU or/and GPU solutions, 
we consider also a different approach based on remote distributed architecture. 
The input data (e.g. set of LORs) is not processed locally but is 
first anonymized, encrypted and transferred to the distant computing centre and then it is processed by 
computing nodes of a grid or a cloud network~\cite{wislicki}.
This scenario is a part of the broader computing concept being developed  
by the CIS \'Swierk Computing Centre.

The medical data management and storage is not a trivial issue and 
one must consider costs of the maintenance of the 
specialized servers and backup systems. Also, because of the sensitive character
of the medical data the security requirements must be fulfilled. 
In addition, it is expected that the overall size of the medical imaging data that needs
to be stored will grow very fast, and the limited storage space of the local computer resources in a
medical unit can  
become an important limitation.

In the proposed computing model, the processing and the storage of medical 
imaging data are moved from hospitals and healthcare facilities
to dedicated computing centres, which provide specialized cloud services.
The hospitals and healthcare centres play a role of clients.   
As a result, the hospitals are freed from the problem related with the medical data storage,
backup procedures, or disk space limitations. 
Also, the centralization provided by the computing centre improves the level
of security
through the application of the system-wise control and automated security upgrades.
Last but not least, the overall cost of examination per patient with respect
to the data processing and further data storage expenses is expected to be much lower than
in the case of the on-site processing. 
The other advantages of the remote processing model are scalability of the solution
and the facility of data-access. 
Among the disadvantages of this approach, one can find the need
of data transfer between the hospital and the computing centre, which amounts of about 1 GB per examination, 
and would increase
the total access time. 
Also, the remote processing requires per se, 
the access to a reliable transfer link.

Currently the work on the remote architecture model is ongoing.  
The case of the J-PET scanner is very well suited  
because of the relatively high amount of data per examination generated, due to the triggerless 
mode of operations and the full 3-D field of view. Among the current issues
we mention: providing the anonymization of the patients images, the development
of the data reconstruction algorithms and the efficient data transfer model.


\section{Summary}

We presented the overview of the data processing scheme developed for the prototype TOF-PET scanner. The proposed solution is optimized in terms of computation time and resources costs.  Currently, our  work is aimed at  integration of all components and preparation of the system for test measurements with a full-scale prototype. The computing model for the remote data processing is being developed.

\section{Acknowledgements}
We acknowledge technical and administrative support by  A. Heczko, M. Kajetanowicz, W. Migda\l, and the financial support by The Polish National Center for Research and Development through grant No. INNOTECH-K1/IN1/64/159174/NCBR/12, and through LIDER grant 274/L-6/14/NCBR/2015, and The Foundation for Polish Science through MPD program and the EU, MSHE Grant No. POIG.02.03.00-161 00-013/09.



\begin{thebibliography}{1}
\bibitem{Karp2008} J. S. Karp et al., 
         Journal of  Nuclear  Medicine 49 (2008) 462.
%
\bibitem{Kardmas2009} D. J. Kardmas et al., 
         Journal of  Nuclear  Medicine 50 (2009) 1315.
\bibitem{patent_pm} P.~Moskal 
Patent No. WO2011008119-A2.
\bibitem{patent_pm2} P.~Moskal 
Patent No. WO2011008118-A2.
\bibitem{NovelDetectorSystems} P. Moskal et al.,
  Bio-Algorithms and Med-Systems 7 (2011) 73; [arXiv:1305.5187 [physics.med-ph]].
\bibitem{NIM15PM} P. Moskal, N.~Zo\'n et al.,
   Nuclear Instruments and Methods in Physics Research Section A 775 (2015) 54.
%
\bibitem{NIM15LR} L. Raczy{\'n}ski et al.,
 Nuclear Instruments and Methods in Physics Research Section A 786 (2015) 105.
\bibitem{NIM14LR} L. Raczy{\'n}ski, P.~Moskal, P.~Kowalski, W.~Wi\'slicki et al.,
 Nuclear Instruments and Methods in Physics Research Section A 764 (2014) 186; [arXiv:1407.8293 [physics.ins-det]].
\bibitem{NIM14PM} P. Moskal, S. Nied\'zwiecki et al.,
  Nuclear Instruments and Methods in Physics Research Section A 764 (2014) 317; [arXiv:1407.7395 [physics.ins-det]].
%

\bibitem{StripPETconcept} P. Moskal et al., 
  Nuclear Medicine Review 15 (2012) C68; [arXiv:1305.5562 [physics.ins-det]].

\bibitem{TOFPETDetector} P. Moskal et al.,
  Nuclear Medicine Review 15 (2012) C81; [arXiv:1305.5559 [physics.ins-det]].


  
\bibitem{x4}
  A. Wieczorek, B. Zgardzi\'nska, B. Jasi\'nska, M. Gorgol, {\it et al.},
  {\em Acta Phys. Polon.} {\bf A127} (2015) 1487.
\bibitem{x5}
  A. Wieczorek, P. Moskal, Sz. Nied\'zwiecki, {\it et al.},
  {\em Nukleonika} {\bf 60} (2015) 777.
\bibitem{x3}
  E. Kubicz, B. Jasi\'nska, B. Zgardzi\'nska, {\it et al.},
  {\em Nukleonika} {\bf 60}, (2015) 749.
\bibitem{x6}
  P. Moskal {\it et al.},  
  {\em Acta Phys. Polon.} {\bf B47} (2016) these proceedings.
  
\bibitem{x7}
  D. Kami\'nska {\it et al.},
  {\em Nukleonika} {\bf 60} (2015) 729.
%
\bibitem{korcyl}  G. Korcyl, P. Moskal et al.,
  Bio-Algorithms and Med-Systems 10 (2014) 37. 
	
\bibitem{korcyl2}  G. Korcyl et al.,
  %
  Acta Phys. Pol. B (2016) these proceedings. 

\bibitem{krzemien_parallel}
  W. Krzemie\'n {\it et al.},
  {\em Nukleonika} {\bf 60} (2015) 745.
\bibitem{palka} M. Pa{\l}ka et al.,
  Bio-Algorithms and Med-Systems 10 (2014) 41; [arXiv:1311.6127 [physics.ins-det]]. 

\bibitem{CStheory1} E. Candes, J. Romberg, T. Tao, 
	IEEE Transaction on Information Theory 52 (2006) 489.

\bibitem{moskal_neha}
  P. Moskal, N. Sharma, M. Silarski  {\it et al.},
  Acta Phys. Pol A127 (2015) 1495-1499.
\bibitem{krzemien} W. Krzemie\'n et al.,
Acta Phys. Polonica A Vol. 127, No. 5 (2015). arXiv:1503.00465 [physics.ins-det].
%
%

\bibitem{boost}  Website: BOOST. Available at: http://www.boost.org/.
\bibitem{root} R. Brun, F. Rademakers  
 Nuclear Instruments and Methods in Physics Research Section A 389 (1997).

\bibitem{shepp} L. A. Shepp, Y. Vardi
IEEE Trans.Med. Imaging, MI-1 No. 2, 113-122 (1982).
\bibitem{slomski} A. S\l omski, Z. Rudy et al., 
Bio-Algorithms and Med-Systems 10,  1-7 (2014). arXiv:1504.06889 [physics.med-ph].
\bibitem{bialas1} P. Bia\l as, J. Kowal, A. Strzelecki  et al., 
Acta Phys. Polon. B Suppl. 6, 1027-1036 (2013).
\bibitem{bialas2} P. Bia\l as, J. Kowal, A. Strzelecki et al., 
Bio-Algorithms and Med-Systems Vol. 10, No. 1, 9-12 (2014).

\bibitem{bialas_GPU}
  P. Bia\l as, J. Kowal, A. Strzelecki  {\it et al.},
  Acta Phys. Pol A127 1500-1504 (2015).
\bibitem{shepp_logan}
L. A. Shepp and B. F. Logan, IEEE Transactions on Nuclear Science, No. 21 (3):21–43 (1974).

\bibitem{wislicki} W. Wi\'slicki et al.,
Bio-Algorithms and Med-Systems. Vol. 10, No. 2, 53 (2014). arXiv:1401.6929 [physics.comp-ph].

%



%
%







\end{thebibliography}
\end{document}